# Search For Hole Mediated Ferromagnetism In Cubic (Ga,Mn)N

M. Sawicki[1], T. Dietl[1], C.T. Foxon[2], S.V. Novikov[2], R.P. Campion[2], K.W. Edmonds[2], K.Y. Wang[2], A.D. Giddings[2], B.L. Gallagher[2]

*1) Institute of Physics, Polish Academy of Sciences, al. Lotników 32/46, Warszawa, Poland.
2) School of Physics and Astronomy, University of Nottingham, Nottingham, United Kingdom*

**Abstract.** Results of magnetisation measurements on p-type zincblende-(Ga,Mn)N are reported. In addition to a small high temperature ferromagnetic signal, we detect ferromagnetic correlation among the remaining Mn ions, which we assign to the onset of hole-mediated ferromagnetism in $Ga_{1-x}Mn_xN$.

## INTRODUCTION

For widespread technological usage of ferromagnetic semiconductors [1], a Curie temperature $T_C$ significantly above 300 K is necessary. In this context, the mean-field p-d Zener model prediction of the room temperature hole-mediated ferromagnetism in (Ga,Mn)N [2] stimulated much interest. However, contrary to (Ga,Mn)As, Mn does not seem to be an efficient acceptor in GaN. Nevertheless, there are numerous observations of room temperature ferromagnetism in (Ga,Mn)N, even in samples that appear to be n-type [3]. The origin of the ferromagnetism in these cases is unresolved but there is little evidence that the ferromagnetic-like behaviour is brought about by carrier mediated coupling. Hole-induced ferromagnetism appears more probable in cubic (Ga,Mn)N, as Mn incorporation is favoured in this phase and since large polar effects, which hinder effective p-type doping, are absent. In fact, we recently demonstrated that this material can be highly *p*-type, with carrier concentrations approaching $10^{18}$ cm$^{-3}$ at room temperature for desirable Mn contents [4]. Here we report on magnetic studies of such a material, which point to the presence of the onset of the hole-mediated ferromagnetism.

## SAMPLES AND EXPERIMENTAL

Cubic (Ga,Mn)N layers, of typical thickness 300 nm, were grown on semi-insulating GaAs (001) substrates by plasma-assisted molecular beam epitaxy using arsenic as a surfactant to initiate the growth of cubic phase material [5]. Films were grown under N-rich conditions at growth temperatures from 450 to 680°C. The Mn concentration in the films was set using the *in-situ* beam monitoring ion gauge, and calibrated *ex situ* by secondary ion mass spectrometry (SIMS). Arsenic concentration in the layers is below 0.5 %, which is known to reduce GaN band gap less than 50 meV [6]. To ensure electrical isolation as well as rule out the possibility of Mn diffusion into the GaAs layer being responsible for the observed electrical and magnetic properties, an undoped cubic GaN (~150 nm thick) followed by a cubic AlN buffer layers (50-150 nm thick) were introduced between the GaAs and cubic (Ga,Mn)N layers in some samples. Hall effect measurements unambiguously reveal that the layers are *p*-type. The nature of relevant acceptor is unknown at present, but we note that layers grown under the same conditions but without Mn are *n*-type. The hole density $p_{Hall}$ generally increases with increasing Mn up to ~4 %, reaching a plateau or decreasing slightly above this value [4]. Since $p_{Hall}$ at 300 K is found to be in the range $10^{17}$ to $10^{18}$ cm$^{-3}$, the doping level is below the Mott critical concentration, and the samples show localisation of the carriers at low temperatures.

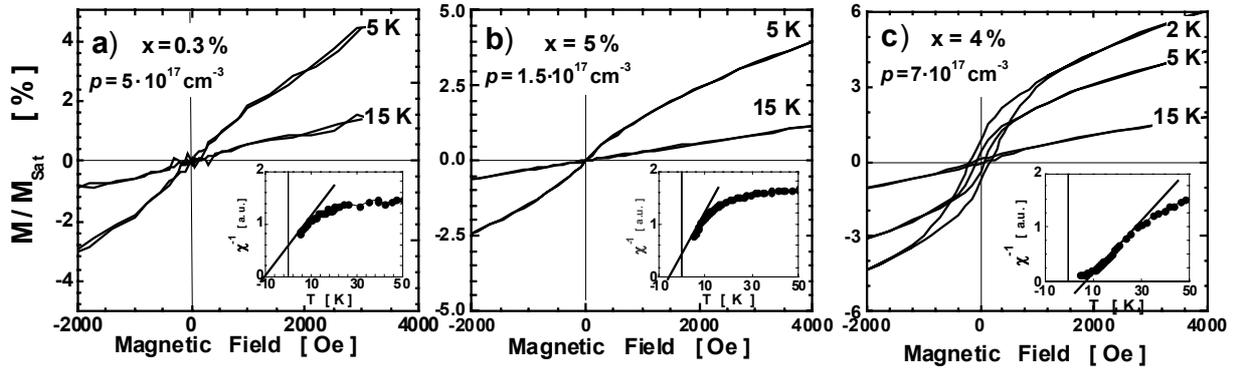

**FIGURE 1.** Hystersis loops at selected temperatures for zb-(Ga,Mn)N with various hole and Mn concentrations (as established by SIMS). The data are plotted relative to the expected saturation for the stated Mn contents.

Magnetic measurements are performed in a SQUID magnetometer, with magnetic fields of up to 4 kOe applied in the plane of the sample. Room temperature measurements evidence the presence of a weak, on average below $0.1\ \mu_B$/Mn, ferromagnetic signal. Similarly to earlier reports, it persists to above 400 K, so MnAs formation due to Mn diffusion to GaAs substrate can be excluded. The origin of this ferromagnetic signal is not presently known, however there are several $Mn_xN_y$ phases with ferromagnetic transition temperature above room temperature which could account for the observed behaviour [7]. Here we concentrate on investigating the majority of Mn that is not involved in the high temperature coupling. The measured magnetisation of the room temperature ferromagnetic phase is nearly temperature-independent below 200 K. We therefore subtract magnetisation curves measured at 50 K from those measured at lower temperatures. Starting with samples having low concentration of either Mn ions (fig. 1a) or holes (fig. 1b) we find the relative magnetisation, $\Delta M(H,T) = M(H,T) - M(H,50\ K)$, to be nearly linear with the external magnetic field, characteristic of a paramagnetic phase. A considerably modified behaviour is observed for samples having higher hole densities (as shown on fig. 1c). There is a much stronger low field curvature and a clear hysteresis is developing on lowering temperature. This is a clear indication of an additional, low temperature magnetic coupling between the Mn ions in this sample. This conjecture is further supported by low temperature $\chi^{-1}(T)$ (plotted in the insets), which changes from a negative value for the first two samples to clearly positive value for the most heavily doped sample. We ascribe this to the presence of holes, which favours a *ferromagnetic* alignment of substitutional Mn at low *T*. We also note that the magnetisation curves of figure 1c are reminiscent of those observed for *p*-type (Zn,Mn)Te single crystals [8]. Even though those samples exhibited insulating behaviour at low *T*, recent inelastic neutron measurements indicated local ferromagnetic ordering mediated by weakly localised holes [9]. However, other mechanisms may also give rise to ferromagnetic ordering at low temperatures in *p*-type materials, including percolation of bound magnetic polarons [10]. Clearly, more work is required to resolve this issue.

This work was supported by EU FENIKS project (EC:G5RD-CT-2001-00535), UK EPSRC (Gr/46465 and GR/S81407), and Polish KBN grant PBZ-KBN-044/P03/2001.